\documentclass[a4paper,11pt]{article}
\usepackage{jinstpub} 
\usepackage{lineno}
\usepackage{algorithm}
\usepackage{algpseudocode}
\usepackage{sidecap}

\usepackage{subcaption}

\title{Challenges and strategies in verification of FastRICH ASIC for the LHCb RICH detector}

\author[a,1]{M.~Lupi,\note{Corresponding author.}}
\author[a]{R.~Ballabriga,}
\author[a]{F.~N.~Bandi,}
\author[a]{G.~Bergamin,}
\author[a]{D.~Ceresa,}
\author[b]{D.~Gascon,}
\author[c]{S.~Gomez,}
\author[a]{J.~Kaplon,}
\author[b]{R.~Manera,}
\author[b]{J.~Mauricio,}
\author[a]{A.~Patern\`o,}
\author[a]{D.~Peninon-Herbaut,}
\author[a]{A.~Pulli,}
\author[a]{S.~Scarf\`i,}
\author[a]{G.~J.~Wegrzyn,}
\author[a]{and K.~Wyllie}
 
\affiliation[a]{CERN, 1, Esplanade des Particules, Meyrin, Switzerland}
\affiliation[b]{ICCUB, Martí i Franquès, 1, Barcelona, Spain}
\affiliation[c]{UPC, Carrer Jordi Girona, 31, Barcelona, Spain}

\emailAdd{matteo.lupi@cern.ch}

\abstract{
The FastRICH ASIC provides high-precision, triggerless readout for the LS3 Enhancements and Upgrades II of the LHCb RICH detector. The demands of continuous data acquisition and varying hit rates across the detector impose unique challenges on the ASIC's design and verification. This work presents the verification strategy for FastRICH, focusing on functional correctness, timing performance, and operational robustness. The methodology includes simulations across occupancy scenarios, validation of timing precision, and stress testing under pile-up and high-rate conditions. Results demonstrate that FastRICH meets its performance requirements over the full range of expected occupancies. Key design and verification challenges specific to triggerless, fast-timing ASICs are discussed, along with lessons learned for future developments.
}


\arxivnumber{2511.03857}

\begin{document}
\maketitle
\flushbottom

\section{Introduction}\label{sec:intro}
The LHCb RICH detector readout electronics and sensors will be replaced in LS3 Enhancements and Upgrades II, respectively~\cite{tdr:lhcb:rich}.
The FastRICH ASIC was designed to provide $25 ps$ timing resolution, combined with continuous readout.
The fine time resolution poses unique challenges to the design and verification.
In particular, clock distribution, jitter, quantization error, and flip-flop metastability resolutions have a major impact on the capability of the verification environment to assess the correctness of the ASIC output.

The common approaches used in verification, based on comparing the predicted with the effective output, do not work.
In this contribution, a novel approach based on approximate scoreboarding is described.
The verification framework was also re-used to run semi-directed tests, in the form of scans, for verifying that the digital processing did not introduce artifacts or performance bottlenecks.

The FastRICH ASIC is briefly introduced in section~\ref{sec:fastrich}; the verification strategy is described in section~\ref{sec:vrf}; finally sections~\ref{sec:tdc} and~\ref{sec:readout} describe the measurement of performance with semi-directed tests.

\section{FastRICH ASIC}\label{sec:fastrich}
The FastRICH ASIC was designed to address the RICH detector requirements.
In particular, the detector required a 16-channel ASIC able to readout different sensors~\cite{tdr:lhcb:rich}.
The ASIC should be able to readout at most one hit per Bunch Crossing (BX) in a configurable window, with a Time-To-Digital (TDC) timing resolution of $25 ps$.
The ASIC should also sustain a per-channel occupancy ranging from $0\%$ to $30\%$, depending on the physical position in the detector.
The ASIC needs to support a continuous readout, be compatible with the lpGBT, and operate in a radiation environment ($TID < 50 kGy$ for LHC run 5).

\begin{figure}[h]
\caption{The FastRICH ASIC is composed of 16 channels, each with a dedicated digital processing. The variable-length data packets are processed, strictly in order, in the packet processor and frame builder. Before being streamed on the serial links, the frames are encapsulated in the Aurora protocol.}
\centering
\includegraphics[width=0.99\textwidth]{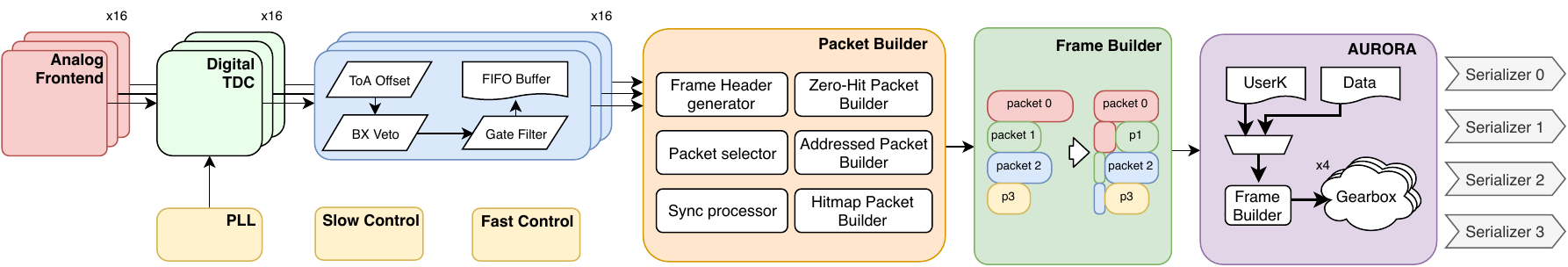}\label{fig:block}
\end{figure}

The FastRICH ASIC, in Figure~\ref{fig:block}, is a 16-channel readout ASIC with multiple configurable analogue front-ends in each channel, capable of resolving multiple hits per BX.
A digital filtering (time gate) functionality allows rejecting hits that occur outside a configurable window inside each BX.
In order to cope with the data rate variability, the ASIC implements a data-driven architecture with zero suppression.
Moreover, the variable-length data format is optimized to minimize the output bandwidth.
The data packets are encapsulated into the Aurora 64b-66b protocol~\cite{aurora}, to ease data recovery in case of Single-Event Effects (SEEs).
The data are transmitted over four configurable serial links ($320 Mbps$ to $1.28 Gbps$).
The ASIC implements selective Triple-Modular Redundancy (TMR) to increase the SEE tolerance: the control logic is completely triplicated (combinatorial and sequential logic, resets, clocks, voters), whereas the data path is not triplicated.

\section{Verification Strategy}\label{sec:vrf}
The main verification goals for the FastRICH ASIC are to check timing correctness and readout correctness: the two goals have to be met at the same time.
The verification strategy implements a coverage-driven constrained-random verification based on the Universal Verification Methodology (UVM)~\cite{ieee:uvm}.
Multiple verification components, developed for other ASICs in the High-Energy Physics community, were re-used~\cite{chips:vip}.
The DUT was verified with the same verification framework at different level of abstraction: RTL, triplicated RTL, and Gate-Level Netlist (GLN), with annotated delays.
Finally, the verification of SEE tolerance is treated as an additional verification goal and it is factored in from the design phase of the verification environment.

In order to verify and validate the required timing resolution, the verification environment was carefully engineered to account, early-on, for the sources of uncertainty.
The clock is provided to the ASIC with $5 ps$ peak-peak jitter, which represents an upper bound of the jitter produced by an lpGBT elink~\cite{lpgbt:specs,lpgbt:meas}; the PLL model also accounts for a realistic amount of jitter, $8 ps$ peak-peak, in line with worst case from analog simulations.
We modeled metastability and metastability random resolution in selected flip-flops in the TDC already in RTL.
This was a key step in verifying the functionality of the fully-digital TDC and validating its performance without the need for time-consuming implementation re-spins.

The protocol correctness is checked in all tests under different conditions.
In particular, it is to be noted that, depending on the channel occupancy, measured in hits per channel per BX, the ASIC can operate between two regimes:
\begin{itemize}
    \item \emph{Low occupancy}. 
    The ASIC mostly transmits empty packets.
    In this regime, the LHC BX and Orbit~\cite{lhc:timing} tracking\footnote{BXID wraps around at every LHC Orbit. Orbit tracking refers to keeping track of which LHC Orbit a specific BXID corresponds to.} are the only pieces of information that are transmitted.
    The verification framework should ensure that the timing information can be correctly tracked and that the latency introduced by the Aurora framing is within acceptable limits.
    No spurious packets should be present in this operating mode.
    \item \emph{High occupancy}.
    The ASIC mostly transmits packets with multiple hits.
    This regime stresses the internal memories and results, first, in back-pressure and, later, in data losses.
    The verification environment should ensure that the data losses do not result in protocol violations and that, once the occupancy drops below an acceptable limit, the data transmission resumes correctly.
\end{itemize}

The test used for verifying the data readout functionality is based on two main phases:
\begin{itemize}
    \item \emph{Preparatory phase}. 
    In this phase, the DUT is reset, randomly configured, and then the PLL and fast control are locked.
    Finally, one hit is injected in each channel and it is readout.
    These hits act as a reference hit, used later on, to calculate the expected time of each hit.
    During SEE verification, faults are injected in the DUT in this phase.
    \item \emph{Repeated data taking phases}.
    During these phases, the occupancy is randomized once for each repetition.
    Random hits are injected with the selected occupancy and they are readout.
    During the SEE verification, faults are injected in these phases.
    It is to be noted that, after each phase leading to data losses (for very high occupancy or SEEs), a subsequent phase with standard occupancy and no SEEs follows.
    The latter allows verifying that the data losses are only transitory and that they do not require a reset of the DUT.
\end{itemize}

Additionally, we also employed formal checks to complement the functional verification framework: static and connectivity checks target at verifying that analog macros are correctly connected to the rest of the design.
For the sake of brevity, these will not be further treated in this contribution.

\subsection{Checker strategies}\label{sec:scb}
Verification of the FastRICH ASIC required novel approaches to functional checking. Standard checker architectures are usually built around a reference model that predicts the expected output frames, coupled with a scoreboard that compares the DUT output frames against these predictions. 
However, this approach is poorly suited to FastRICH, where precise prediction of readout data is intrinsically difficult due to its extremely fine time resolution and asynchronous TDC operation.

At a binning of 25~ps, even small uncertainties in clock phase distribution across channels can determine whether a hit is assigned to one TDC bin or another.
In gate-level simulations with annotated timing, the exact phase of the sampling clock at each channel can vary, making a cycle-accurate reference model infeasible.
Furthermore, the asynchronous nature of the TDC introduces an inherent $\pm$1 bin ambiguity in digital simulations, caused by random metastability resolution in sequential elements.  

Another complication arises from the possibility of input hits being lost due to time gate filtering or TDC dead times.
To precisely determine whether a particular input hit is filtered, would require a timing-accurate behavioral model of the TDC, which is extremely challenging to implement due to its asynchronous nature and variable delays introduced in gate-level simulations.  This further limits the feasibility of conventional prediction-based checkers.  

To overcome this, the verification environment adopted an unconventional abstraction for scoreboarding. 
Instead of predicting readout frames directly, the DUT's output was first decoded into individual channel hits.
These observed hits were then reconciled with the input stimulus hits using a classification algorithm.
The algorithm categorized hits as matched, mismatched, ghost, or lost, enabling systematic evaluation of data integrity, without requiring deterministic prediction of exact bin placement.
Crucially, the algorithm incorporated tolerance for ambiguous cases, such as hits near shutter edges or those subject to $\pm$1 bin assignment uncertainty.

The classification algorithm is summarized in Algorithm~\ref{alg:classification}.
It processes each observed hit against a queue of expected hits, applying a hierarchy of matching rules and filtering checks until a final classification is reached.

This abstraction, shifting the comparison domain from output frames to input hits, enabled scalable and robust checking across the wide dynamic range of occupancies foreseen for FastRICH.
The strategy, not only simplified verification under high-precision timing conditions, but also provided comprehensive coverage of rare failure modes, such as simultaneous multi-hit events or near-saturation behavior.
By addressing the challenges of asynchronous TDC behavior and high-precision timing resolution, this checker methodology represents a generalizable solution for verifying future high-precision ASICs, where traditional frame-level reference models become impractical.

\begin{algorithm}[H]
\caption{Hit Classification Algorithm}\label{alg:classification}
\small
\scriptsize
\begin{algorithmic}[1]
\Require observed hit $o$, expected queue $Q$, last observed hit $o_{last}$, vetoed BXIDs $V$
\Ensure classification of $o$ and corresponding expected events, $e$
\While{$o$ is unclassified}
  \If{$Q$ is empty}
    \State classify $o$ as \textsc{Ghost}
  \Else
    \State $e \gets Q.pop\_front()$
    \If{ToA$(e)$ matches ToA$(o)$}
      \If{ToT$(e) \neq$ ToT$(o)$}
        \State classify both as \textsc{ToT Mismatch}
      \ElsIf{$o$ outside gate}
        \State classify $o$ as \textsc{Ghost}, $e$ as \textsc{Gate Filtered}
      \ElsIf{$o_{last}$ exists and BXID$(o_{last}) =$ BXID$(o)$}
        \State classify $o$ as \textsc{Ghost}, $e$ as \textsc{Bero Filtered}
      \ElsIf{BXID$(o) \in V$}
        \State classify $o$ as \textsc{Ghost}, $e$ as \textsc{BX Vetoed}
      \Else
        \State classify $o,e$ as \textsc{Match}
      \EndIf
    \ElsIf{$e$ filtered by TDC/gate/BX veto}
      \State classify $e$ accordingly
    \ElsIf{$Q$ is empty}
      \State classify $o,e$ as \textsc{ToA Mismatch}
    \Else
      \State classify $e$ as \textsc{Lost}
      \If{ToA$(o)$ not close to ToA$(e)$ or ToA$(Q.front)$}
        \State classify $o$ as \textsc{Ghost}
      \EndIf
    \EndIf
    \State save classification of $e$
  \EndIf
\EndWhile
\If{$o$ is \textsc{Hitmap Only}}
  \State classify $o$ as unexpected
\EndIf
\State save $o$ as $o_{last}$
\end{algorithmic}
\end{algorithm}

\section{Verification of TDC timing performance}\label{sec:tdc}
The random test described in section~\ref{sec:vrf} was re-used to run semi-directed tests in the form of scans.
These scans allowed looking at the ASIC behavior at a higher level of abstraction.
Two examples developed for the FastRICH verification are the Time of Arrival (ToA) scan and Test Pulse (TP) scan.
The first studies the TDC linearity, the latter the TP delivery network.
In both cases, one hit (or TP) was injected in each BX (and each channel) each with an increasing ToA.
The whole $25 ns$ of a BX are swept with steps of $1 ps$.
These hits (or TPs) are readout and different tests are repeated with different time gate configurations.
These scans allowed verifying that the TDC and clock distribution do not introduce artifacts in the data.

\section{Verification of readout performance}\label{sec:readout}
Two additional types of scans were executed to study readout efficiency and readout latency \emph{vs} data occupancy.
These two scans allow verifying that the readout behaves consistently and predictably at different data occupancies.
For both scans, the test was repeated in different scenarios (number of output lanes enabled, data rate, etc.).
The data from these scans were used to identify scenarios where the ASIC behavior was well within the specifications but could be improved, providing important feedback to the designers.

\begin{figure}[h]
\caption{TDC dead time scan. 
The requirement of being able to readout two hits in two consecutive BX translates to a TDC deadtime smaller than $18.87 ns$, i.e. a $25ns$, a full BX, minus the maximum time gate duration, $6.25 ns$.
This corresponds to two hits in subsequent BXs, where the first hit is at the end of the time gate and the second hit is at the beginning of the time gate, in the subsequent BX. 
Here, it can be observed that all the channels, in different colors, are well within the required limit and that they all behave consistently.}
\centering
\includegraphics[clip, trim=0cm 0cm 0cm 0cm, width=0.9\textwidth]{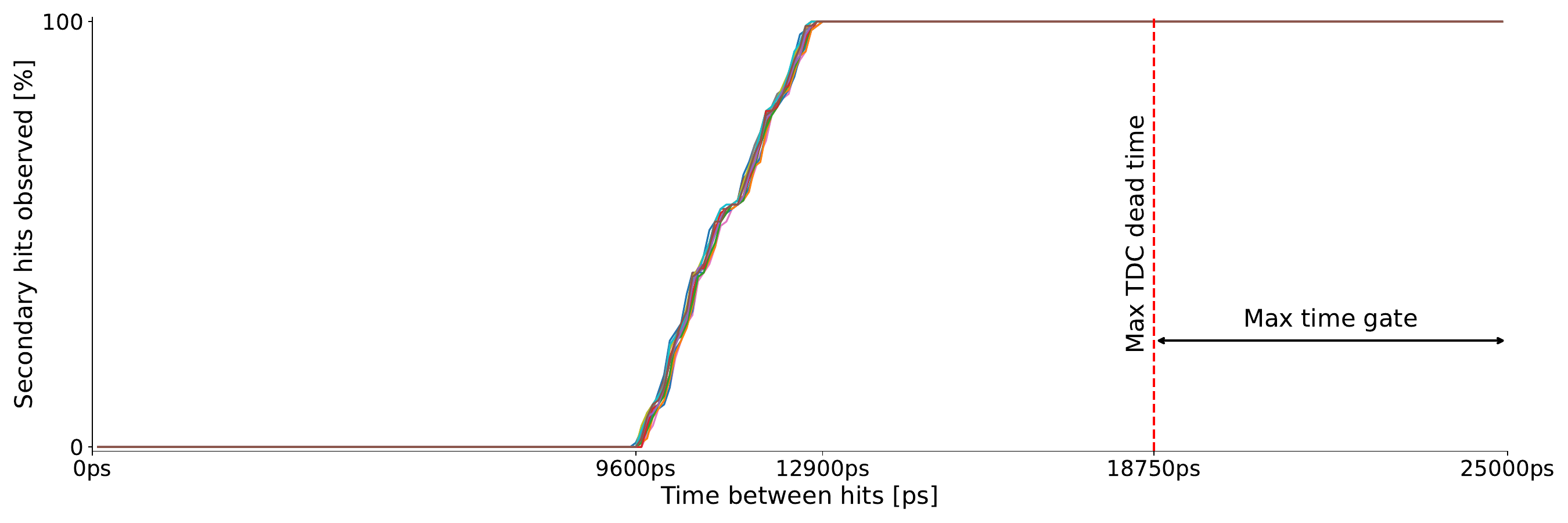}\label{fig:tdc}
\end{figure}

Another important study executed was the validation of the TDC dead time.
The digital TDC, should not limit the performance of the sensor or the analog front-end.
In this scan, two hits are injected in each BX at increasing distance in time $\delta_t$.
The percentage of secondary hits is recorded vs $\delta_t$.
An example, extracted from the gate-level simulation at typical corner is shown in Figure~\ref{fig:tdc}.


\vspace{-.5cm}

\section{Conclusions}\label{sec:end}
The verification strategy of the FastRICH ASIC was presented.
The approach based on modeling metastability in key sequential elements in the digital TDC, together with an innovative scoreboarding technique, allowed verifying and validating the ASIC performance already at the RTL stage.
This saved a considerable amount of time in the iterations at the GLN verification.
The presented approach re-used the tests written for the constrained-random verification in order to study the behavior of the TDC at a higher level of abstraction, ensuring that no artifacts of the digital processing chain are present in the output data.

FastRICH was submitted for fabrication in February 2025 and the first packaged prototypes were received in July 2025.
The ASIC is currently undergoing an in-depth characterization in hardware and a beam test is taking place at the time of writing.


\end{document}